\documentclass{article}

\usepackage{arxiv}

\usepackage[numbers]{natbib} 
\usepackage{tabularx,longtable,multirow,subfigure,caption}
\usepackage{subcaption} 
\usepackage{url} 
\usepackage[english]{babel}
\usepackage{amsmath}
\usepackage{graphicx}
\usepackage{dsfont}
\usepackage{epstopdf} 
\usepackage{backref} 
\usepackage{array}
\usepackage{latexsym}
\usepackage[pdftex,pagebackref,hypertexnames=false]{hyperref} 
\usepackage{algorithm}
\usepackage{algorithmic}
\usepackage[all]{hypcap}
\usepackage{subcaption}
\usepackage{graphicx}

\title{From Deep Filtering to Deep Econometrics}

\date{14 September 2023}

\author{ \href{https://orcid.org/0009-0004-6902-1595}{\includegraphics[scale=0.06]{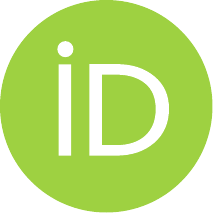}\hspace{1mm}Robert Stok}\\
	Department of Computing\\
	Imperial College London\\
	South Kensington Campus\\
    London SW7 2AZ\\
	\texttt{rs1720@ic.ac.uk} \\
	\And
	\href{https://orcid.org/0000-0000-0000-0000}{\includegraphics[scale=0.06]{orcid.pdf}\hspace{1mm}Paul Bilokon}\\
	Department of Mathematics\\
	Imperial College London\\
	South Kensington Campus\\
    London SW7 2AZ\\
	\texttt{paul.bilokon@imperial.ac.uk}\\
}

\renewcommand{\shorttitle}{From Deep Filtering to Deep Econometrics}

\hypersetup{
pdftitle={From Deep Filtering to Deep Econometrics},
pdfauthor={Robert Stok, Paul Bilokon},
pdfkeywords={stochastic filtering, particle filtering, deep filtering, econometrics, deep econometrics},
}

\begin{document}
\maketitle

\graphicspath{ {images/} }

\begin{abstract}
Calculating true volatility is an essential task for option pricing and risk management. However, it is made difficult by market microstructure noise. Particle filtering has been proposed to solve this problem as it favorable statistical properties, but relies on assumptions about underlying market dynamics. Machine learning methods have also been proposed but lack interpretability, and often lag in performance. In this paper we implement the SV-PF-RNN: a hybrid neural network and particle filter architecture. Our SV-PF-RNN is designed specifically with stochastic volatility estimation in mind. We then show that it can improve on the performance of a basic particle filter.
\end{abstract}

\section{Introduction}

\subsection{Motivations}
Volatility forecasting plays a crucial role in finance and risk management, particularly in options pricing \cite{Black1973} \cite{Ball1994} and risk assessment \cite{Shephard2009}. While volatility itself is not directly observable, it can be estimated by analyzing related time series, such as price shocks of correlated financial assets \cite{Tsay2010}.

Traditionally, volatility estimation has relied on ARCH/GARCH models, which have demonstrated effectiveness \cite{Hansen2005} \cite{Bollerslev1986}. However, these models have limitations in that they do not account for exogenous variables and fail to capture certain characteristics of volatility, such as long-memory, jumps, and leverage effects \cite{Zivot2009} \cite{Chan2016}.

Recently, researchers have explored the use of machine learning methods for volatility forecasting \cite{Ge2022}. Hybrid approaches combining machine learning and traditional statistical methods have shown promise \cite{Christensen2021}, with recurrent neural network (RNN) models delivering notable results due to their ability to capture time series dependencies \cite{Nguyen2022}.

However, a common criticism of machine learning methods is their lack of interpretability \cite{Christensen2021}. Moreover, as functional rather than probabilistic models, they do not provide a comprehensive understanding of the distribution of estimated volatility, which is crucial for robust risk modeling.

Another popular approach to modeling volatility is through stochastic volatility models, which employ Monte Carlo methods to approximate the time-varying probability distribution \cite{Shephard2005}, accounting for changes in underlying stock prices. Particle filters are commonly used to approximate this distribution \cite{Pitt1999} \cite{Malik2011}. 

\subsection{Objectives}
In this paper, our objective is to adapt the PF-RNN model proposed by Karkus et al. \cite{Karkus2018}. to the task of volatility filtering. We call this the SV-PF-RNN. This architecture is a fully differentiable particle filter, providing the advantages of inference while generating a set of particles that approximates the likelihood of volatility.

Subsequently, we aim to demonstrate that our model can achieve or surpass the performance of a particle filter in estimating volatility from generated sequences. By generating volatility sequences and corresponding sets of returns using a stochastic volatility model, we train our hybrid model and evaluate its performance against the particle filter.

Finally, we undertake a thorough analysis of the limitations inherent in the SV-PF-RNN model. We identify specific challenges and propose potential remedies to address these shortcomings, while also highlighting future avenues for development on the SV-PF-RNN model.

\subsection{Contributions}
Our paper makes the following key contributions:

\begin{enumerate}
  \item Implementation of a hybrid particle filter and recurrent neural network model for volatility filtering.
  \item Introduction of randomness to the input of the neural network to allow it to learn non-linear noise distributions in its transition function.
  \item Method for accelerating training by individually pretraining the neural networks.
  \item Novel loss function for learning stochastic models incorporating randomness as an input to a neural network.
  \item Comparison of the trained model against a standard particle filter using generated data.
\end{enumerate}

By addressing the limitations of existing approaches and introducing our hybrid architecture, we aim to advance the field of volatility forecasting and contribute to improved risk modeling techniques.

\section{Preliminaries}

\subsection{Financial Preliminaries}
In this section we will go over the financial background knowledge that is relevant to our task at hand. We will first review options and then the pricing of those options to contextualize the importance of volatility calculation. We then go over two common methods of modelling volatility. Finally we give a brief introduction to particle filters and recurrent neural networks, as these two concepts will feature heavily in our implementation.

\subsubsection{Options}
An option is a derivative that gives the buyer of the option (but not the obligation) to buy or sell the underlying financial asset at a particular price \cite{Hull2017}. This price is called the strike price. The contract is voided at a date known as the ‘expiry’, and can be exercised either on the expiry or anytime up to and including the expiry depending on if it is a European or American option respectively. When an option is sold the seller takes on the risk of losing money if the stock price goes up or down, as shown on the graphs below. Note that a call option gives the buyer the right to purchase the stock and a put option gives them the right to sell it. Other terminology that will often be brought up is ‘short’ and ‘long’ positions. A ‘long’ position means that you have bought an option, whereas a ‘short’ position is the equivalent of selling an option.

\begin{figure}[tb]
\centering
\includegraphics[width = 0.8\textwidth]{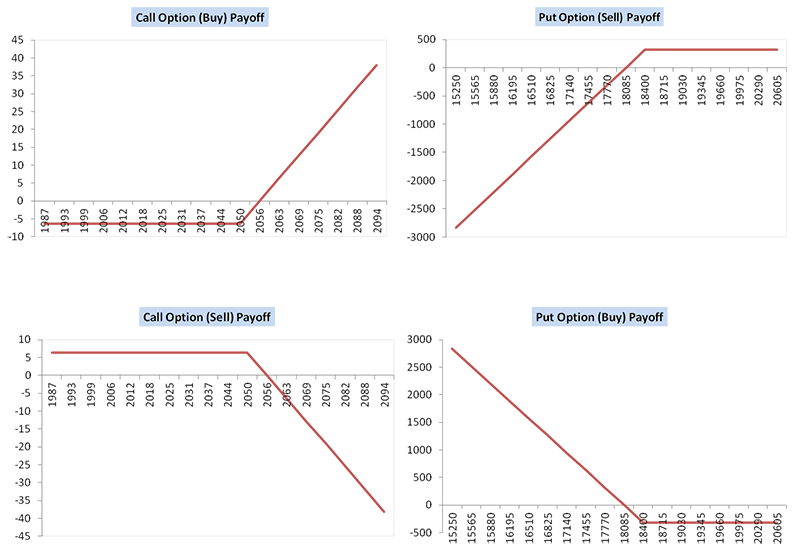}
\caption{Graphs showing the payoffs for put and call options}
\label{fig:logo}
\end{figure}

For the seller (i.e. if you hold a short position) you are taking on risk, as changes in the price of the financial asset could cost you money. The price of the option, known as the ‘premium’, is the payment for this risk, and thus stocks which are more likely to exhibit price movements are more expensive. The relative size of the price movements in a financial asset are known as its ‘volatility’. If a participant in the market is able to more accurately predict volatility then they can make money by trading volatility. This can be achieved by buying or selling a combination or put and call options to create a ‘straddle’ \cite{Hull2017}. As the graphs below show this combination will earn money either if the stock does not change price, or exhibits large changes in price, regardless of the direction of change. Intuitively, this is one of the major motivators behind trying to better estimate volatility.

\begin{figure}[tb]
\centering
\includegraphics[width = 0.8\textwidth]{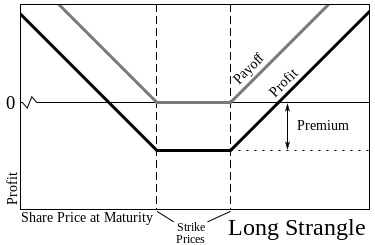}
\caption{Graph showing the payoff for a Long Strangle from \cite{Gtix})}
\label{fig:logo}
\end{figure}

\subsubsection{Black-Scholes Model for Pricing Options}
In their 1973 article Fischer Black and Myron Scholes proposed the Black-Scholes model as a method of pricing options \cite{Black1973}. You can begin by describing the generalized movement of a stock price with the following stochastic differential equation: 

\begin{equation} \label{stochdiffeq}
    \frac{d S}{S}=\mu d t+\sigma d W
\end{equation}

Here \( S \) is the price of the stock at time \( t \), \( \mu \) is a drift term and \( \sigma \) is the volatility. Let us also imagine we have a twice differentiable equation for the price of an option, \( f(S, t) \). Applying Ito’s lemma we get the following equation:

\begin{equation} \label{itoslemma}
    d f=\left(\mu S \frac{\partial f}{\partial S}+\frac{\partial f}{\partial t}+\frac{1}{2} \frac{\partial^2 f}{\partial S^2} \sigma^2 S^2\right) d t+\frac{\partial f}{\partial S} \sigma S d z
\end{equation}

Black and Scholes’ key insight was to use a ‘hedged’ portfolio, consisting of an option and \( \frac{\partial f}{\partial S} \)shares of the underlying asset, to price the option. We can describe the value of this portfolio \( \Pi \) as follows:

\begin{equation} \label{hedgedportfolio}
    \Pi=-f+\frac{\partial f}{\partial S} S
\end{equation}

By substituting in our PDEs for the value of a call option and the price evolution of the underlying asset we can reach the following final parabolic PDE describing the value of an option \( f \):

\begin{equation} \label{optionvalue}
    \frac{\partial f}{\partial t}+r S \frac{\partial f}{\partial S}+\frac{1}{2} \frac{\partial^2 f}{\partial x^2} \sigma^2 S^2 - r f = 0
\end{equation}

Where \( r \) is the risk free rate of return.  It is often possible to solve for an exact solution given the type of option and certain boundary conditions \cite{Black1973}.

\subsection{Volatility Modelling}
\subsubsection{ARCH/GARCH Models}
Auto Regressive Conditional Heteroskedastic (ARCH) models are a popular approach to modeling volatility that use lagged values of the underlying asset’s price innovations. This was introduced in Robert Engle’s seminal 1982 paper \cite{Engle1982}, where it was used to model the volatility of inflation in the United Kingdom. Let us assume we have an asset who’s returns follow a random walk. In particular, the returns \( y_t \) are drawn from a normal distribution with variance \( \sigma_{t}^{2} \), where \( rho_t \) can be considered the volatility at time \( t \). 

\begin{equation} \label{arch1}
y_t=\sigma_t \epsilon_t
\end{equation}

Engle realized that volatility is affected by previous returns, which leads to phenomena such as volatility clustering (shown in the diagram below). Therefore, an \( ARCH(p) \) model takes the squares of the previous \( p \) returns to determine the volatility for the current timestep. The equation for the volatility at time \( t \) is shown below.

\begin{equation} \label{arch2}
\sigma_t^2=\alpha_0+\alpha_1 y_{t-1}^2
\end{equation}

\begin{figure}[tb]
\centering
\includegraphics[width = 0.8\textwidth]{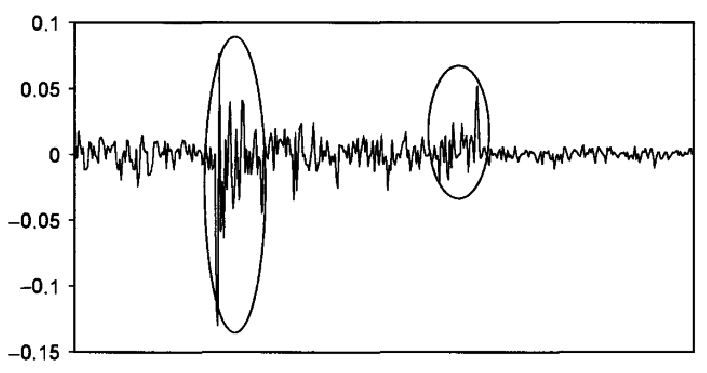}
\caption{Graph showing an example of volatility clustering from \cite{Alexander2000}}
\label{fig:logo}
\end{figure}

In 1986 Bollerslev introduced the GARCH model \cite{Bollerslev1986}, a generalized form of the ARCH model. Although the ARCH model works, it fails to capture some stylized facts of volatility. The first is that negative returns will cause volatility to increase more than positive returns of the same magnitude. This is known as the asymmetric leverage effect \cite{Campbell1992}. The second is that periods of high or low volatility tend to persist \cite{Baur2018}.

To account for this the \( GARCH(p, q) \) model extends the standard \( ARCH(p) \) model to include the squares of the volatility in the previous \( q \) timestamps. The equation is shown below for a \( GARCH(p, q) \) model

\begin{equation} \label{arch3}
    \sigma_t^2 = \omega + \sum_{i=1}^p \alpha_i \epsilon_{t-i}^2 + \sum_{j=1}^q \beta_j \sigma_{t-j}^2
\end{equation}

\subsubsection{Stochastic Volatility}
Stochastic volatility is one of the main fields of volatility modeling used to deal with the issue of time varying volatility. In basic, non-stochastic option pricing models such as Black-Scholes, a constant volatility is assumed, hence the variance of the underlying asset’s movements is considered fixed and we can use MLE to estimate this variance for the entire series. In stochastic volatility we replace this with a function of time \( \sigma_t \) that models the variance of the \( S_t \) term of the brownian motion equation. Intuitively \( \sigma_t \) is the time varying volatility of the underlying financial asset. We can assume \( \sigma_t \) follows some type of brownian motion, however, exact form depends on the specific model being used. 

Several authors separately developed on the ideas of stochastic volatility, but a few models made notable progress \cite{Shephard2005}. One of the earliest models was Taylor’s model \cite{Taylor1987} \cite{Taylor1994}, which demonstrated the effects of volatility clustering (described in the ARCH/GARCH modeling section). The Taylor model starts by modeling the log price returns of the underlying asset \( s_t \) as a gaussian process using a noise term \( \varepsilon_{t}^{S} \sim \mathcal{N}(0, 1) \) multiplied by the volatility \( \sigma_t \). \( \mu \) is a drift term, which cancels out the risk free rate of return.

\begin{equation} \label{logreturns}
y_t = log(S_{t}) - log(S_{t-1})
\end{equation}
\begin{equation} \label{volatilityandreturns}
y_t = \mu + \sigma_t \varepsilon_{t}^{S}
\end{equation}

\( \sigma_t \) is modeled using the following equations, where \( h_t \) is a gaussian process with a non-zero mean, and \( \varepsilon_{t}^{H} \sim \mathcal{N}(0, \tau^2) \).

\begin{equation} \label{volatilityexp}
\sigma_t = e^{h_t/2}
\end{equation}
\begin{equation} \label{stochvoleq}
h_t = \mu + \phi\left ( h_{t-1} - \mu \right ) + \varepsilon_{t}^{H}
\end{equation}

Another feature of some stochastic volatility models is their ability to incorporate leverage effects. By negative correlating the noise variables \( \varepsilon_{t}^{S} \) and \( \varepsilon_{t}^{S} \) Neilson was able to incorporate the asymmetry of the effect of returns on volatility \cite{Shephard1998}. There is also empirical evidence of this effect in stochastic volatility models applied to real world data \cite{Xu2010}.

\subsection{Particle Filters}
Particle filters are a class of Monte Carlo algorithms that can be used to solve state estimation problems with non-linear belief state spaces \cite{Gordon1993}. They work by approximating the belief space using a set of ‘particles’, which each represent a potential state and its likelihood. These are updated at each timestep according to the dynamics of the system and then reweighted using external observations.

For a better understanding let us formally describe the problem. We have a series of unknown states \( x_{1}, x_{2},...,x_{t} \), and a series of observations \( z_{1}, z_{2},...,z_{t} \). An observation at time \( t \) is related to a state at time \( t \) by the probability distribution \( p(z_t \mid x_t) \), which is known as the observation likelihood function. Similarly, a state in time \( t \) is related to previous states by \( p(x_t \mid p_{t-1},...,p_0) \), which can be simplified to \( p(x_t \mid p_{t-1}) \) for markov chain processes. This is known as the state transition function. Our aim is to calculate \( p(x_t \mid z_{t-1},...,z_0) \). 

Since calculating this posterior distribution is often intractable, particle filters aim to approximate it through a set of weighted particles \( \mathcal{P}_t=\{h_{t}^{i},w_{t}^{i}\}_{i=0}^{n} \), where \( h_t \) is a belief state at time t, and \( w_t \) is the weight of the particle, which can be thought of as its likelihood. Instead of using transition functions and likelihood functions that are probability distributions we can write them as equations with noise to represent our uncertainty, and then sample from this:

\begin{equation} \label{pftransfunc}
    p_{t}^{i} \sim p(p_t \mid p_{t-1}^{i})
\end{equation}
\begin{equation} \label{pfobsfunc}
    w_{t}^{i} \propto p(y_t \mid p_{t}^{i})w_{t-1}^{i} 
\end{equation}

At each step in time we update each particle by applying the state transition equation, which gives us an approximation of the prior distribution \( p(x_t \mid p_{t-1}) \). We then update the weights on each of the particles using the known likelihood function \( p(z_t \mid x_t) \) to get an approximation of the posterior distribution \( p(x_t \mid z_t) \). A common issue is particle degeneracy, where only a few particles have meaningful weights. To solve this a resampling step is introduced in algorithms such as Sequential Importance Resampling (SIR). This also allows the particle filter to focus on approximating the most important/promising regions.

The full particle filter algorithm is written below:

\begin{algorithm}
\caption{Calculate $y = x^n$}
\begin{algorithmic} 
\REQUIRE $\mathcal{P}_{t-1} = \{h_{t-1}^{i},w_{t-1}^{i}\}_{i=0}^{n} \text{ and observation } y_t$
\FOR{$i \leftarrow 0...n$}
\STATE $\hat{h}_{t}^{i} \leftarrow P_{trans}(h_{t-1}^{i})$
\STATE $w_{t}^{i} \leftarrow w_{t-1}^{i} * P_{obs}(x_{t}^{i}, y_t)$
\ENDFOR
\FOR{$i \leftarrow 0...n$}
\STATE $\hat{w}_{t}^{i} \leftarrow \frac{w_{t}^{i}}{\sum_{i=1}^n w_{t}^{i}}$
\ENDFOR
\STATE $\textup{generate sample index } S \textup{ from discrete probability distribution created from } \hat{w}_{t}^{i}$
\FOR{$i \leftarrow 0...n$}
\STATE $j \leftarrow S^i$
\STATE $h_{t}^{i} \leftarrow \hat{h}_{t}^{j}$
\STATE $w_{t}^{i} := \frac{1}{n}$
\ENDFOR
\RETURN $\{h_{t}^{i},w_{t}^{i}\}_{i=0}^{n}$
\end{algorithmic}
\end{algorithm}

There are different methods of getting a final estimate from the set of particles, but the most common is simply to take a weighted mean:

\begin{equation} \label{weightedmean}
    \hat{x_t} = \frac{1}{n}\sum_{i=0}^{n}h_{i}^{t}w_{i}^{t}
\end{equation}

\subsection{Machine Learning}
\subsubsection{Recurrent Neural Networks}
One issue with traditional neural networks is that they are not good at handling sequences of information \cite{Goodfellow2016}, especially long or variable length sequences. To solve this issue John Hopfield introduced the first recurrent neural network in his 1982 paper \cite{Hopfield1982}. The idea behind recurrent neural networks is that a sequence of data is passed into the network one time step at a time. The network then feeds part of its output back into itself to help inform the decision at the next timestamp. These networks can either produce a prediction at each timestep, or produce one or more predictions at the end.

\begin{figure}[h!]
    \centering
    \includegraphics[width = 0.8 \textwidth]{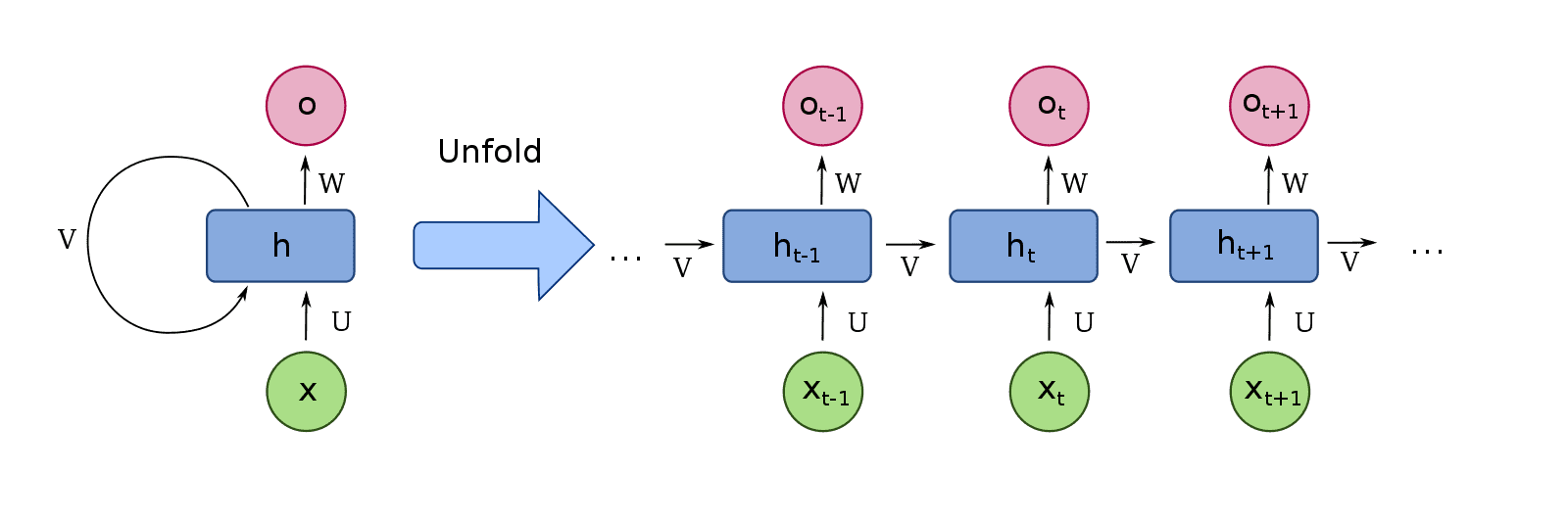}
    \caption{Diagram illustrating how an RNN works from \cite{Adaloglou2020}}
    \label{fig:rnn}
\end{figure}

These networks are trained using back propagation through time \cite{Mozer1995}. Intuitively this can be thought of as ‘unrolling’ the network, so that there is a copy of the network for each timestep, connected by the weights that make up the feedback loop of the network. Through this connection errors in the later outputs can be used to train weights based on earlier inputs. This is illustrated in figure \ref{fig:rnn}.

\begin{figure}[h!]
    \centering
    \includegraphics[width = 0.6 \textwidth]{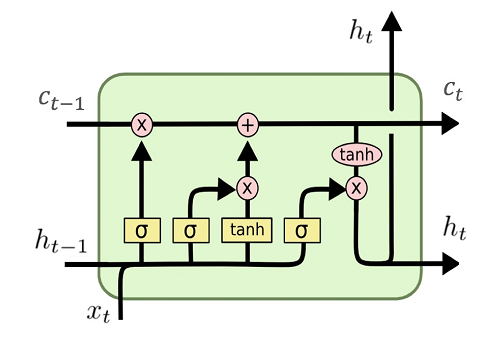}
    \caption{Diagram of an LSTM from \cite{Rahuljha2020}}
    \label{fig:LSTM}
\end{figure}

One of the more popular types of RNN is an LSTM, which was introduced by Hochreiter and Schmidhuber in \cite{Hochreiter1997}. It deals with the issue of long term dependencies by introducing a cell state, as well as a hidden state. The cell state can be thought of as the long term memory of the cell, and at each timestep part of it is ‘forgotten’ and part of it is ‘updated’ with the network's hidden state. On the other hand, the hidden state is similar to the hidden state of a regular RNN. Fig \ref{fig:LSTM} is a diagram of an LSTM along with the set of equations that govern its workings.

\begin{equation}
    i_t =\sigma\left(x_t U^i+h_{t-1} W^i\right)
\end{equation}
\begin{equation}
    f_t =\sigma\left(x_t U^f+h_{t-1} W^f\right)
\end{equation}
\begin{equation}
    o_t =\sigma\left(x_t U^o+h_{t-1} W^o\right)
\end{equation}
\begin{equation}
    \tilde{C}_t =\tanh \left(x_t U^g+h_{t-1} W^g\right)
\end{equation}
\begin{equation}
    C_t =\sigma\left(f_t * C_{t-1}+i_t * \tilde{C}_t\right)
\end{equation}
\begin{equation}
    h_t =\tanh \left(C_t\right) * o_t
\end{equation}

\section{Background}
In this chapter in split into three main sections. In the two sections we will go over previous work that has been done on volatility prediction. In particular we separately look at work that has focused on machine learning methods, and work that has focused on particle filtering methods. In the second section we look at work that has been done on combining particle filters and neural networks, although this work has not yet been applied to the field of econometrics.

\subsection{Volatility Prediction}
\subsubsection{Neural Network Approaches}
There is a large body of work that suggests combining traditional statistical methods with machine learning methods can yield better results than either of the two in isolation. One way to do this is to assume that volatility has both a linear and non-linear component and then separately model these with statistical and neural network approaches. In his 2003 paper Peter Zhang \cite{Zhang2003} uses an ARIMA model to forecast volatility and then trains a neural network to predict on the residuals. Although the hybrid model manages to outperform the ARIMA model no information criterion analysis is done. Christensen, Siggaard and Veliyev \cite{Christensen2021} compare an extended HAR model to three neural networks of increasing size, on predicting 1 day out volatility for 29 different stocks on the DOW jones index. They find that the neural networks outperform the HAR model with up to an 11.8\% reduction in the MSE on certain financial instruments. They also address the common criticism that ML models are not interpretable by running an ALE analysis, allowing them to see which factors contribute most heavily to the network’s prediction.

\subsubsection{Recurrent Neural Networks}
Recurrent neural networks are a popular choice for many time series forecasting tasks due to their ability to infer patterns in long sequences of data \cite{Ge2022}. However, purely RNN based methods have varying results, and are often beaten by statistical methods in common time series prediction tasks \cite{Makridakis2018}. Yang \cite{Liu2019} compared GARCH, v-SVR and LSTM models on 3 day out volatility prediction. He found that although LSTMs outperformed the GARCH model, they had comparable performance to v-SVR. Jia and Young compared deep neural networks and LSTMs against GARCH and ARCH models \cite{Jia2021}. While the LSTMs outperformed the ARCH/GARCH models the difference was not significant and care was not taken to find the optimal hyperparameters of the ARCH/GARCH models. In both these approaches the authors use only the series of financial returns as input to the networks.

A far more successful class of algorithms combines statistical methods with RNN methods. Gustavo et al. \cite{DiGiorgi2023} and Yan et al. \cite{Hu2020} both propose LSTM and bidirectional LSTM (BiLSTM) models that use GARCH model forecasts as their inputs. In both cases the authors found that the hybrid models outperformed traditional GARCH models by a significant margin. Gustavo et al. also compare the performance of LSTMs and BiLSTMs with and without the GARCH forecasts. In their paper they give each of the models the price innovations of copper as well as a set of other explanatory variables and perform 2 week out copper price prediction. They found that the models that were given GARCH price forecasts outperformed the models that were not. It is also worth noting that, while hard to compare to other results due to the uncommon dataset, they achieve state of the art performance, potentially due to the use of exogenous explanatory variables. 

Another approach taken by Nguygen et al. \cite{Nguyen2022} introduces a novel architecture called an SR-SV model. Their motivation is to better capture long-memory and non-linear autodependence phenomena in volatility forecasting. In their paper they combine an SRU with a traditional stochastic volatility model as shown in eq. \ref{SRU1} to \ref{SRU4}. Here \( \beta_1 \) is the non-linear component term. An SRU is a type of RNN that allows a vector of summary statistics \( h_t \) to move through the network using a moving average. The equations that govern it are summarized below. They evaluate their network on three different simulated volatility datasets. Datasets 2 and 3 incorporate long-memory and non-linear autodependence into their simulation models. Although the SR-SV has comparable performance on the first model to SV modelling, it outperforms on the second two volatility models.

\begin{equation} \label{SRU1}
    z_t=\beta_0+\beta_1 \operatorname{SRU}\left(\eta_{t-1}, z_{t-1}, h_{t-1}\right)+\phi z_{t-1}+\epsilon_t^\eta
\end{equation}
\begin{equation} \label{SRU2}
    r_t =\Psi\left(W_h h_{t-1}+b_r\right)
\end{equation}
\begin{equation} \label{SRU3}
    \varphi_t  =\Psi\left(W_r r_t+W_x x_t+b_{\varphi}\right)
\end{equation}
\begin{equation} \label{SRU4}
    h_t^{\left(\alpha_j\right)} =\alpha_j h_{t-1}^{\left(\alpha_j\right)}+\left(1-\alpha_j\right) \varphi_t, j=1, \ldots, m ; h_t=\left(h_t^{\left(\alpha_1\right)}, \cdots, h_t^{\left(\alpha_m\right)}\right)^{\top}
\end{equation}

\begin{figure}[h!]
\centering
\includegraphics[width = 0.8\textwidth]{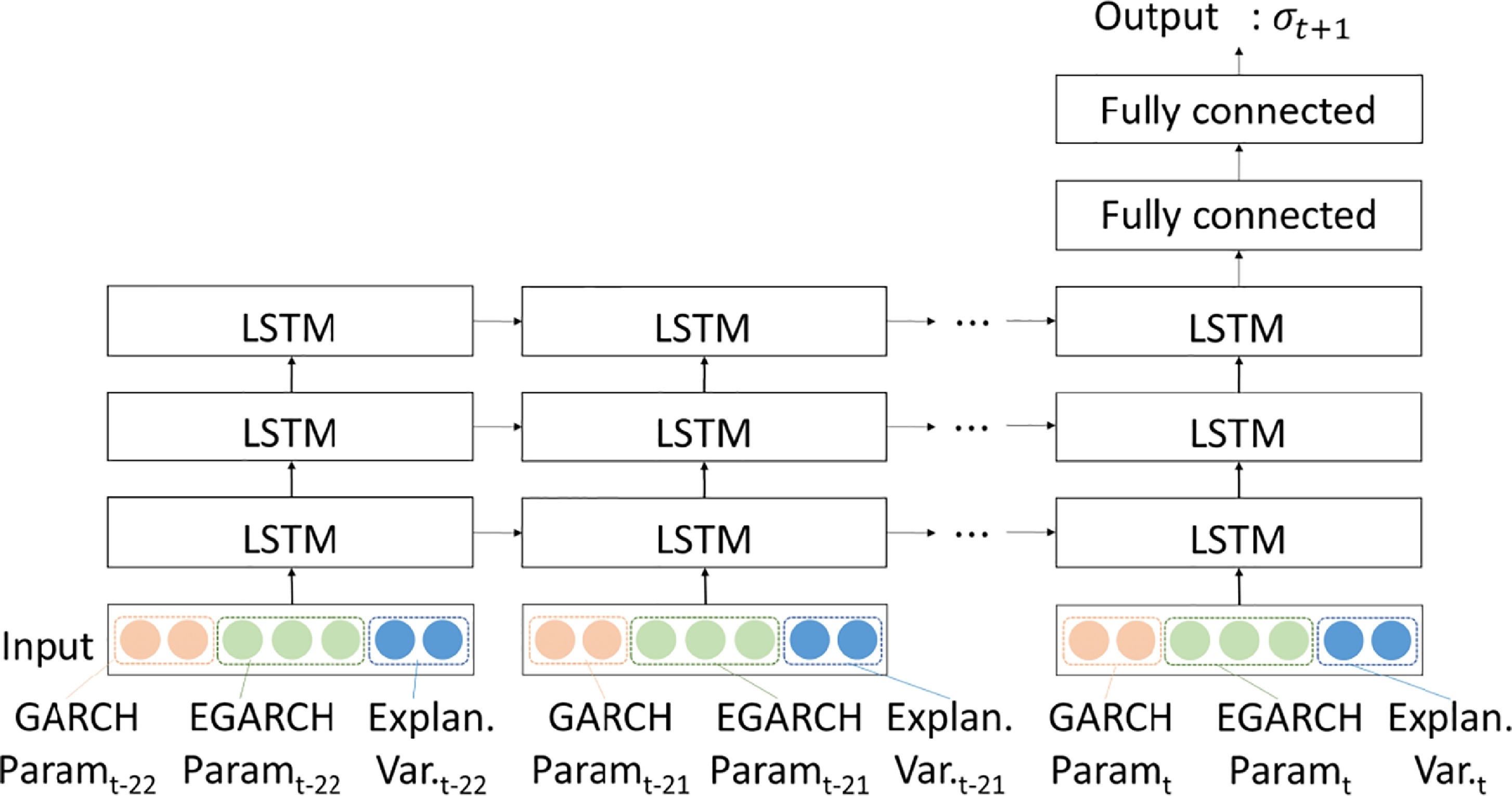}
\caption{Diagram of Kim and Won's hybrid LSTM model from \cite{Kim2018}}
\label{fig:kim2018}
\end{figure}

One of the issues with RNNs is that they need a lot of training data before they begin to outperform other methods. Many of the previous methods assume fixed parameters and so the model may not generalize well to all financial assets. Kim and Won \cite{Kim2018} propose an LSTM that takes the parameters of traditional statistical techniques, such as GARCH and EGARCH, as one of its inputs. These parameters are estimated using MLE, and the model can now be trained on time series with many different parameters. They also included additional explanatory variables such as the price of gold and oil, the CB interest rate and the KTB interest rate. A diagram of this model is shown in fig \ref{fig:kim2018}. In the study they used their model to perform 1 day out volatility predictions for the KOPSI 200 index. They concluded that neural networks have 2 advantages in volatility prediction. The first is that they are able to mitigate the weaknesses of different GARCH-type models through the neural networks. The second is that neural networks are able to utilize econometric information such as gold prices to improve prediction, whereas in standard models these are assumed to just be stationary.

\subsubsection{Convolutional Neural Networks}
Convolutional Neural Networks (CNNs) are most often used for image related tasks due to their spatial pattern recognition capabilities \cite{Krizhevsky2012} \cite{Goodfellow2016}. Although RNNs are more commonly used due to their superior sequential modeling capabilities in time series forecasting tasks, there has been some work in the field of using convolutional neural networks for volatility forecasting. 

J. Doering et al used CNNs to predict stock price and price volatility using data from the London stock exchange \cite{Doering2017}. However instead of using just the price as an input, they used matrix representations of the order book, trades, orders and deletions (of bids/asks) of each stock.  In particular, each column in the matrix represents a particular time window, and each row represents a price point. In their results it can be seen that CNN was able to extract meaningful features and perform successful forecasting, likely due to the additional information. Interestingly the model performed significantly better in the task of volatility prediction compared to the task of price prediction.

\begin{figure}[h!]
\centering
\includegraphics[width = 0.8\textwidth]{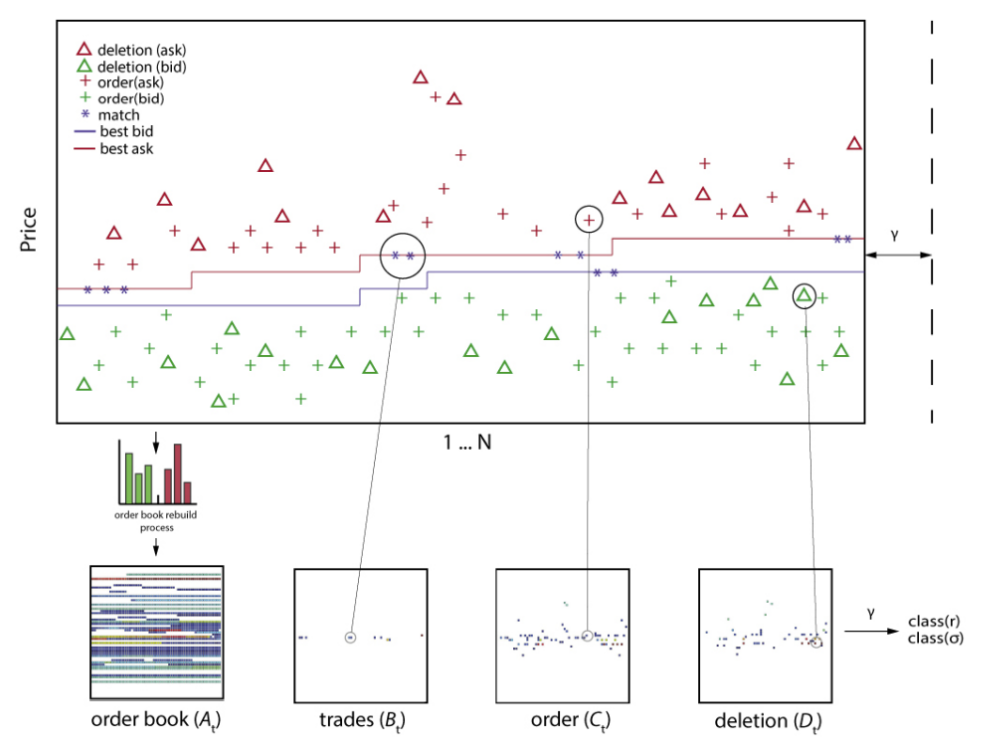}
\caption{Diagram showing how options data is convert to the inputs for the CNN in the Doering et al. paper}
\label{fig:logo}
\end{figure}

\subsection{Particle Filtering}
\subsubsection{Particle Filtering for Parameter Estimation}
Monte Carlo methods have long been employed for the parameter estimation of stochastic volatility models via MLE \cite{Sandmann1998}. Malik and Pitt showed that a particle filter could also be employed to perform likelihood inference on stochastic volatility models and therefore optimize parameters \cite{Pitt2009}. The aim is to estimate the likelihood:

\begin{equation} \label{pfparamest1}
    \log L(\theta)=\log f\left(y_{1, \ldots, .} y_T \mid \theta\right)=\sum_{t=1}^T \log f\left(y_{t+1} \mid \theta ; Y_t\right)
\end{equation}

We can approximate \( f\left(y_{t+1} \mid \theta ; Y_t\right) \) by taking our previous samples/particles from \( f\left(h_{t} \mid Y_t ; \theta\right) \), then sampling from the transition density function \( f\left(h_{t+1} \mid h_{t} ; \theta\right) \), and then exploiting the relationship below:

\begin{equation} \label{pfparamest2}
    f\left(y_{t+1} \mid \theta ; Y_t\right)=\int f\left(y_{t+1} \mid h_{t+1} ; \theta\right) f\left(h_{t+1} \mid Y_t ; \theta\right) d h_{t+1}
\end{equation}

\subsubsection{Particle Filtering for Volatility Estimation}
Particle filters have also been used extensively to directly estimate volatility using on-line filtering. The most common method for building a simple SIR particle filter implementation, also known as a bootstrap filter, for a particular stochastic volatility model, requires two things:

\begin{enumerate}
  \item The definition of a transition density function - this can be derived from the discretized version of the volatility dynamics of the stochastic volatility model in question
  \item The definition of an observation likelihood function - this is derived from the equations relating the estimated volatility to the size of the returns; normally a noise parameter is used at this stage, the shape of which heavily determines the shape of the observation likelihood function
\end{enumerate}

Malik and Pitt build particle filters for four stochastic volatility models: SV, SVL, SVLJ and SV-GARCH \cite{Pitt2014}. They show their derivations of these two functions, including their newly introduced SV-GARCH model. They then evaluate the models on real world data, using log-likelihood as their main criteria. Pitt and Shephard observed that particle filters suffer from two main issues. Firstly, when there is an outlier, the weight distribution is uneven and so a large number of particles is required to draw a distribution close to the empirical sampling density. Secondly, the tails of the distribution \( p \left( y_t \mid x_t \right) \) are often poorly approximated due to a lack of particles. To solve this they implement the auxiliary particle filter (APF) \cite{Pitt1999}. Their key innovation was to introduce an auxiliary variable \( k \), which is then used to draw samples from \( x_{t+1}^j, k^j \sim g\left(x_{t+1}, k \mid Y_{t+1}\right) \). Afterwards the weights are readjusted using the following equation:

\begin{equation} \label{apfreweighting}
    w_{t+1}^j=\frac{f\left(y_{t+1} \mid x_{t+1}^j\right) f\left(x_{t+1}^j \mid x_t^{k^j}\right)}{g\left(x_{t+1}^j, k^j \mid Y_{t+1}\right)}w_t^j
\end{equation}

\( g(.) \) can be designed to make the weights more even. In their paper Malik and Pitt apply this general algorithm to both an ARCH model and a stochastic volatility model. Song et al. also apply the auxiliary particle filter to model stochastic volatility with jumps \cite{Song2014}. They observed that because the tail ends of the observation function are poorly estimated by a SIR particle filter, jumps in volatility are often poorly approximated. To solve this they designed their transition function so that it produces \( k_\lambda=\max \{1,[\lambda \cdot m]\} \) particles with jumps, where \( \lambda \) is the probability of a jump, and \( m \) is the total number of particles. The effect of this change can be seen in the diagram below from their research paper. Although the APF does not perfectly model the distribution, there are at least a good number of particles within the high probability region of the post-jump distribution.

\begin{figure}[h!]
\centering
\includegraphics[width = 0.6\textwidth]{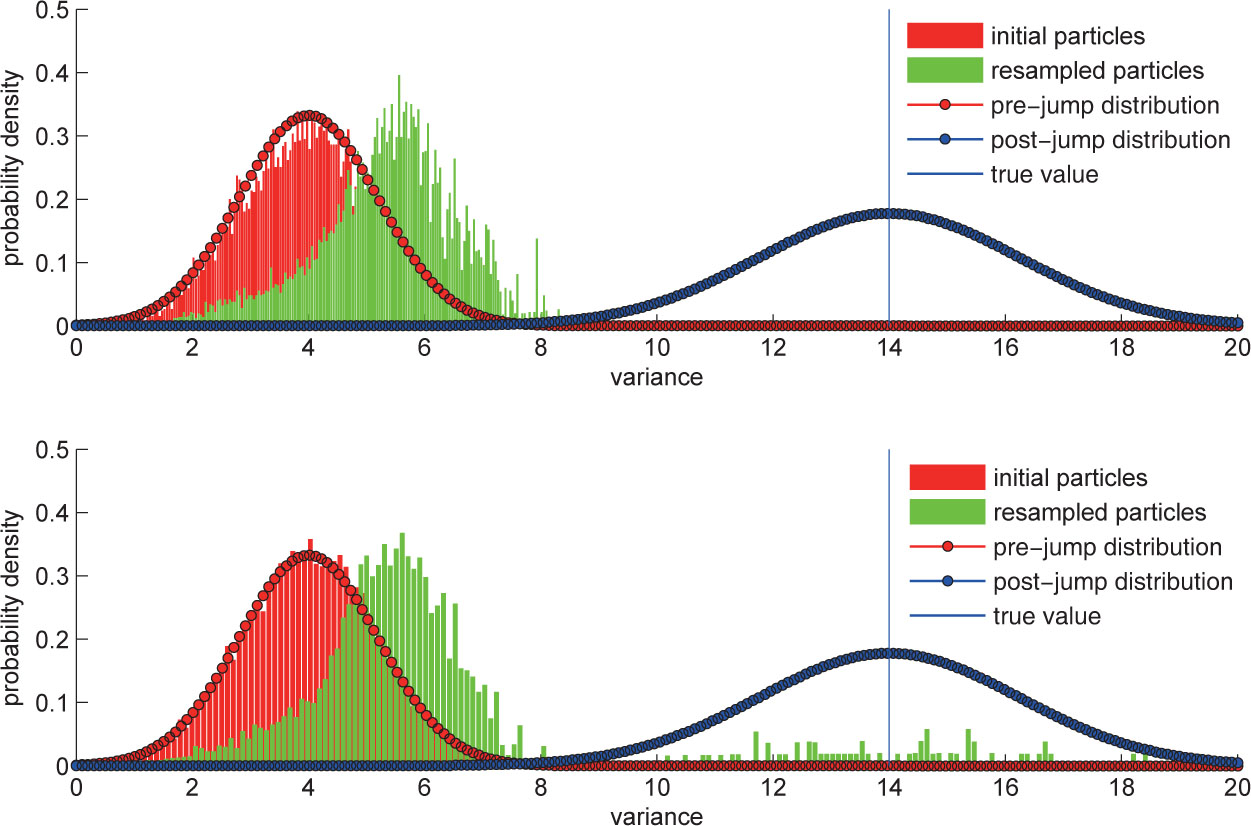}
\caption{Diagram from \cite{Song2014} showing the difference in the estimation capabilities of the vanilla particle filter (top) and the auxiliary particle filter (bottom)}
\label{fig:logo}
\end{figure}

\subsection{Particle Filter Recurrent Neural Networks}
Although particle filters are a good algorithm for approximating the current state in a non-linear state space, care is needed to create good observation likelihood and state transition functions. In \cite{Karkus2018} Karkus, Hsu and Lee introduced the particle filter network (PF-Net): an end to end differentiable implementation of the particle filter. The PF-Net was then used in a 2D localization and visual odometry task. Their work was then further extended by X. Ma et al. in \cite{Ma2020}, who improved upon the basic particle filter network by implementing it as an LSTM and GRU. They used the network for several tasks, but focused on localization within a small generated maze.

Intuitively a PF-Net or PF-RNN is a type of recurrent neural network that has, not just one hidden state \( h_t \), but K hidden belief states with an associated weight \( H_t = \{(h_{t}^{i}, w_{t}^{i})\}_{i=0}^{K} \). These belief states can be thought of as the particles. However, unlike particles they do not necessarily need to directly represent the system’s state but can be a latent representation of it. In a regular RNN the hidden state goes through a deterministic update. however in a PF-RNN they instead go through a stochastic bayesian update \( h_t^i = f_{trans}\left(h_{t-1}^{i}, u_t, \xi_{t}^{i}\right) \) where \( u_t \) is the control vector and \( \xi_{t}^{i}i \) is a noise term. The weights are then updated using a learned observation likelihood function \( w_t^i=f_{obs}\left(x_t, h_t^i\right) w_{t-1}^i \), where \( x_t \) is the observation. In both papers these functions are represented as neural networks with learnable parameters. 

\begin{figure}[h!]
\centering
\includegraphics[width = 0.8\textwidth]{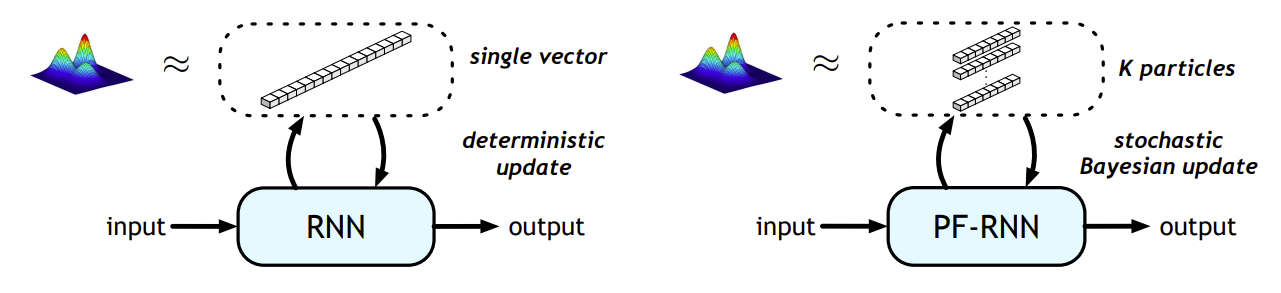}
\caption{Diagram the basic difference between an RNN and a PF-RNN from \cite{Ma2020}}
\label{fig:logo}
\end{figure}

\begin{minipage}{\textwidth}
One of the issues with building a PF-RNN is that resampling is not differentiable. To fix this Karkus et al. introduce ‘soft resampling’. Instead of sampling from the real distribution \( p(i) = w_{t}^{i} \) we sample from another distribution \( q(i) = \alpha p(i) + (1 - \alpha) (\frac{1}{K}) \). This is a combination of the desired distribution \( p(i) \) and a normal distribution, and gives non-zero gradients when \( \alpha \) is larger than 0. The new weights are then calculated using the importance sampling formula:

\begin{equation} \label{importancesampling}
    {w’}_{t}^{i} = \frac{p(k)}{q(k)}
\end{equation}

Ma et al. also introduce a new loss function. They note that the goal of a particle filter is not just to provide an accurate estimate of the exact state, but also to approximate the true probability distribution of the belief state. To do this they optimize an ELBO loss that uses sampled particles:

\begin{equation} \label{ELBO-loss}
L_{\mathrm{ELBO}}(\theta)=-\sum_{t \in \mathcal{O}} \log \frac{1}{K} \sum_{i=1}^K p\left(y_t \mid \tau_{1: t}^i, x_{1: t}, \theta\right)
\end{equation}

Where \( \tau_{i}^{1:t} \) gives the history of samples chosen and random noise inputted into a particle. The value of \( p\left(y_t \mid \tau_{1: t}^i, x_{1: t}, \theta\right) \) is approximated by assuming the distribution is a gaussian with mean 0 and variance 1. They combine this loss with traditional mean square error loss, weighting it with the parameter \( \beta \):

\begin{equation} 
L(\theta) = L_{MSE}(\theta) + \beta L_{ELBO}(\theta)
\end{equation}

\end{minipage}

\section{Contribution}
\subsection{Overview}
Our goal is to predict the volatility of a financial asset over time using a series of observations of its underlying price movements. We can assume we have corresponding pairs of sequences to use as training data. In particular, we have a sequence of observations \( X_1...X_t \) and the sequence of corresponding unobservable volatilities \( Y_1...Y_t \), and our goal is to estimate the volatility at each timestep \( \hat{Y}_1...\hat{Y}_t \). 

\begin{figure}[h!]
\centering
\includegraphics[width = 0.4\textwidth]{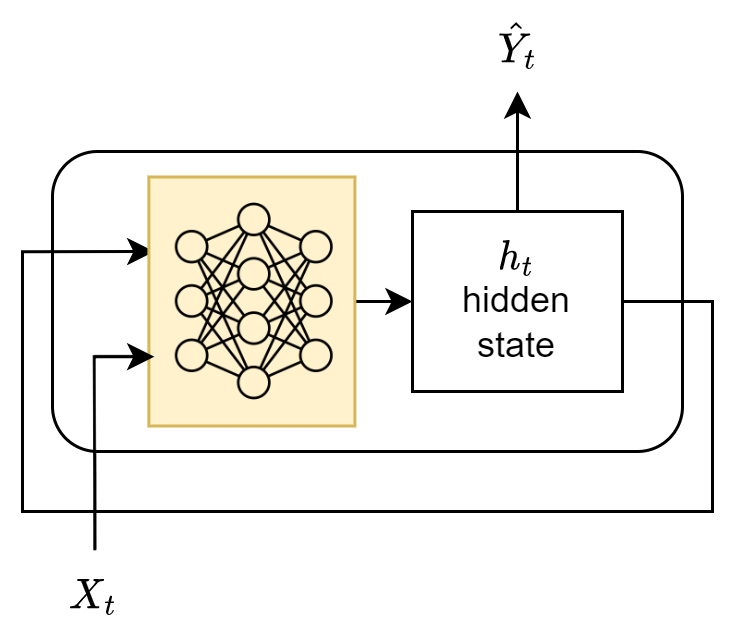}
\caption{Standard RNN implementation}
\label{fig:standard-rnn-impl}
\end{figure}

An approach using an RNN would involve creating a network that takes the price change \( X_t \), and a previous hidden state \( h_{t-1} \), and then outputs the predicted volatility \( Y_t \), and the updated hidden state \( h_t \). The hidden state \( h_t \) and the predicted volatility \( Y_t \) do not necessarily need to be the same, but can be depending on the approach. The network could then be trained using the sequences of data. A diagram of this approach is shown in fig \ref{fig:standard-rnn-impl}.

A particle filter approach requires the definition of a state transition equation \( f_{trans}(p) \), where p is a particle representing the potential state, and the observation likelihood function \( f_{obs}(p, x) \), which given a price movement observation \( x \), returns the likelihood of that particle. The particle's state could just be the predicted volatility, or we can store more information such as the previous \( n \) volatilities if we want to use a model such as a GARCH model. The observation could also be generalized to the previous \( m \) observations for the same reason. Our transition equation and likelihood function would be based on one of the volatility models we have written about in our background section. 

One idea is to adapt the PF-RNN structure to the task of volatility forecasting. We can extend the RNN so that instead of a single hidden state \( h_t \), it has several hidden states, \( \{p_{t}^{i}\}_{i=0}^{n} \) that each represent a belief state, whether that’s just the current volatility estimate or some hidden representation. We can think of these hidden states as particles, and along with the belief state we store the weight associated with each particle, so we have \( \{(p_{t}^{i}, w_{t}^{i})\}_{i=0}^{n} \). The transition function and observation likelihood function are now represented by neural networks which we can train. At each stage the RNN first updates each of the particles, and then uses the current observation and new set of particles to update each of the particles’ weights. Finally the particles are resampled using a differentiable form of resampling called soft resampling. We can take the mean of this new set of particles as the observation.

\begin{figure}[tb]
\centering
\includegraphics[width = 0.8\textwidth]{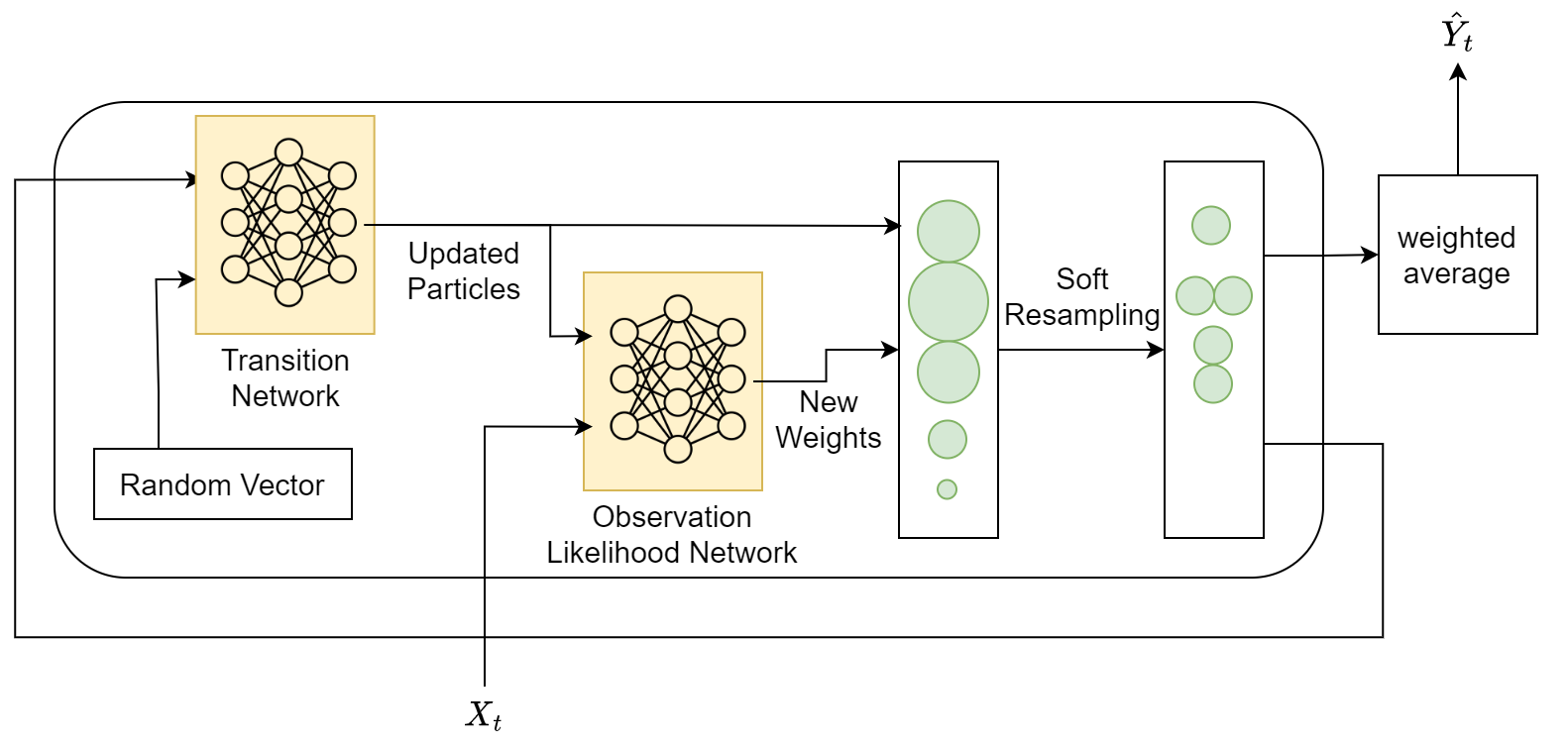}
\caption{Diagram of our SV-PF-RNN (credit to David McDonald from The Noun Project for the Neural Network Icon)}
\label{fig:our-model}
\end{figure}

This is the idea behind the basic SV-PF-RNN that we will be developing in this section. We introduce new architecture to better predict stochastic volatility. We also add randomness so that the transition function can be fully approximated. To help the model reach similar performance to a particle filter we pretrain its networks to approximate a particle filter. We then make several modifications to the loss function to help train the network despite the inherent randomness of volatility calculation, and the issues this creates with low probability outcomes in a SV-PF-RNN. Finally we train our network to show that it can outperform the standard particle filter.

\subsection{Taylor SV-PF-RNN}
\subsubsection{The Taylor Stochastic Volatility Model}
To evaluate the performance of our SV-PF-RNN we will generate data using the relatively simple stochastic volatility model introduced by Taylor \cite{Taylor1987}. The equations for the innovation and the volatility prediction are:

\begin{equation} \label{taylorSV1}
    \sigma_t=\mu+\phi(\sigma_{t-1}-\mu)+\zeta_\sigma
\end{equation}
\begin{equation} \label{taylorSV2}
    x_t=\sigma^2\zeta_x \\
\end{equation}

where \( \zeta_\sigma \sim \mathcal{N}(0, \tau) \) and \( \zeta_x \sim \mathcal{N}(0, 1) \). Note that the model takes three parameters, \( \mu \), \( \phi \), and \( \tau \). We will assume these are fixed. We chose to start with this model as it is simple and will allow us to see if our SV-PF-RNN is able to estimate basic stochastic volatility. 

\subsubsection{Particle Filter Implementation}
Based on this we can define the state transition equation and observation likelihood function for a particle filter implementation. \( \varepsilon \) is a noise variable where \( \varepsilon \sim \mathcal{N}(0, \tau) \). Note that the parameters of the volatility generation are fixed and known, so the state consists only of the predicted volatility. 

\begin{equation}
    f_{obs}(p, x) = \frac{1}{\sqrt{2\pi p^2}}e^{-\frac{x^2}{2p^2}}
\end{equation}

\begin{equation}
    f_{trans}(p, \varepsilon ) = \mu + \phi(p-\mu) + \varepsilon
\end{equation}

\subsubsection{Overview of the SV-PF-RNN}
Our SV-PF-RNN maintains a set of \( n \) particles and associated weights, \( \{(p_{t}^{i}, w_{t}^{i})\}_{i=0}^{n} \), where each particle is a single scalar representing the estimated volatility at that point in time. At each timestep we input a single observation \( X_t \), and then update this belief state using the observation and the previous belief state, and finally output a single estimate of the volatility for that timestep, \( \hat{Y}_t \), based on the belief state.

In our SV-PF-RNN we use two neural networks to represent our transition function and observation likelihood function. We replicate our transition function, \( p_t=f_{trans}(p_{t-1}) \), which can be approximated with a neural network. One issue with this is that the original transition function contains the process noise variable \( \varepsilon \). In Ma and Karkus et al 2020 \cite{Ma2020} the authors add randomness after the transition function, \(p_t=f_{trans}(p_{t-1}) + \varepsilon \), where \( \varepsilon \) has a fixed variance and mean. However in several volatility models the noise is non-linearly used in the transition stage. To fix this we include this noise as an input, and so our transition equation is represented by  \(p_t=f_{trans}(p_{t-1}, \varepsilon ) \), where \( \varepsilon \) is a vector of random numbers drawn from a normal distribution with variance 1 and mean 0. We will discuss the implications of including a random input to the training process later on.

Our observation likelihood function is also represented by a neural network that takes as input the particle and the price observation, and outputs the likelihood, \( w_t=f_{obs}(p_t, x_t) \). Finally after updating and normalizing all our new weights we then resample the particles using soft resampling. To produce our estimated volatility \( \hat{Y}_t \) we simply take the average of all of the particles. Since we resampled in the previous step there is no need to weight each estimate. 

\subsubsection{Pretraining the Networks}
\begin{figure}
\centering
\begin{minipage}{.5\textwidth}
  \centering
  \includegraphics[width=1\linewidth]{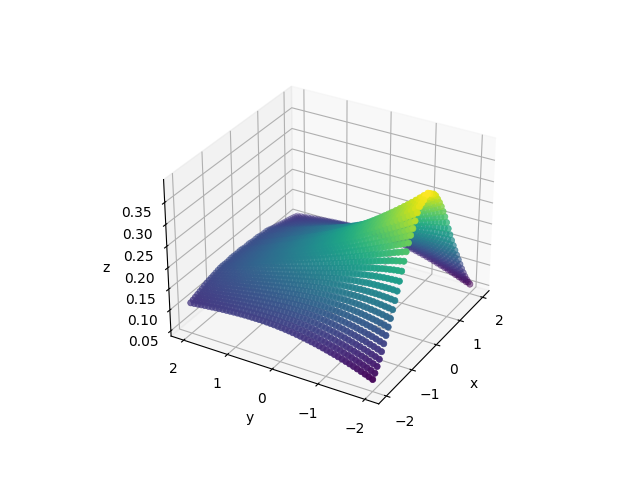}
  \captionof{figure}{Observation likelihood function}
  \label{fig:obs-func}
\end{minipage}%
\begin{minipage}{.5\textwidth}
  \centering
  \includegraphics[width=1\linewidth]{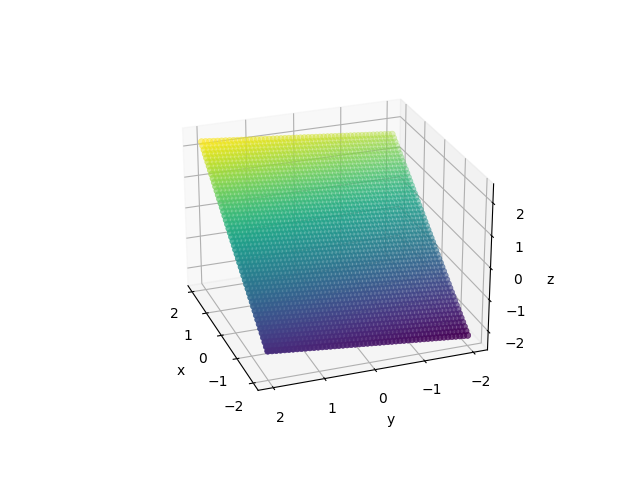}
  \captionof{figure}{Transition function}
  \label{fig:trans-func}
\end{minipage}
\end{figure}

To speed up the training process we first train the network to replicate our particle filter implementation. Neural networks can be thought of as nonlinear function approximators \cite{Hornik1989}. Hence, we can individually train the observation likelihood and transition function networks to approximate their equivalent functions in the particle filter implementation. We can do this by generating input, output data pairs for each of these functions and training the networks on this. We then load the weights into the full model. We have plotted what this looks like for our neural network in figures \ref{fig:obs-func} and \ref{fig:trans-func}.

In our evaluation we show that the performance of this particle filter approximator network is similar to that of the basic particle filter. However, while training we only produced input data without a certain numerical range. In practice values outside this range can sometimes be found, and so the network’s behavior for these values can be unstable.

\subsubsection{Loss Function}
\begin{figure}[tb]
\centering
\includegraphics[width = 0.8\textwidth]{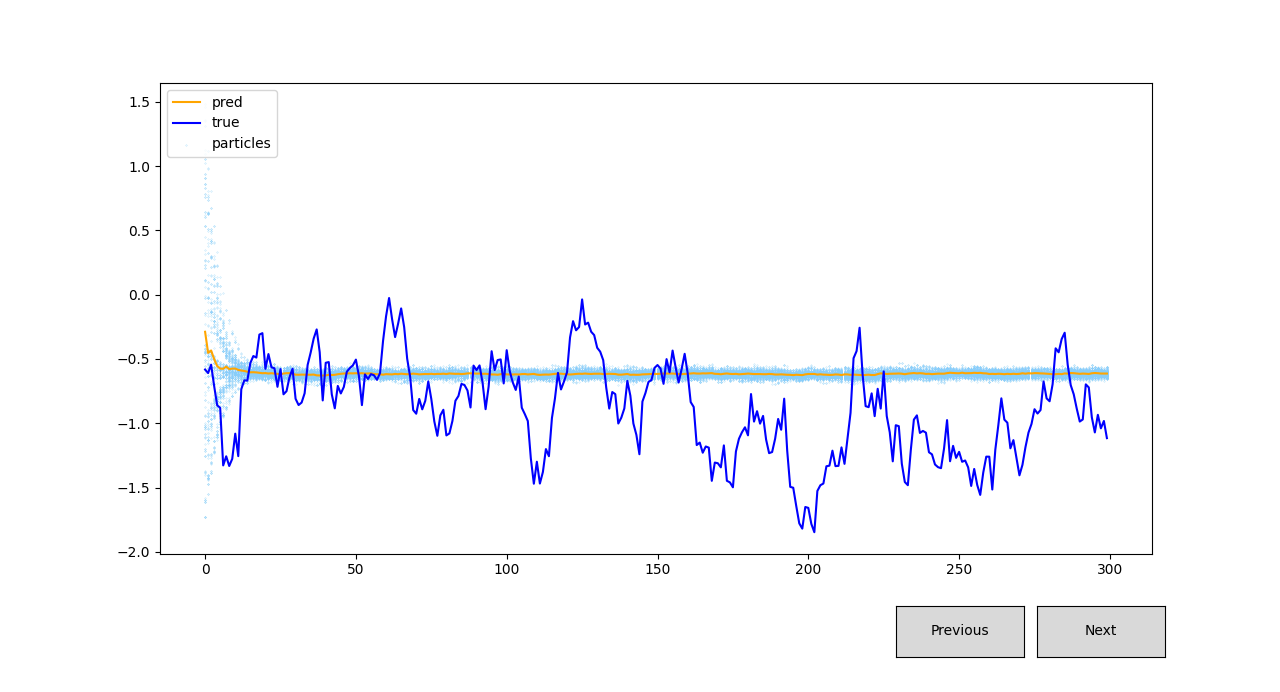}
\caption{The mean convergence issue after training can be seen here; despite the particles starting with a large spread, they soon all converge to value of around \( -0.5 \)}
\label{fig:mean-convergence-issue}
\end{figure}

One of the large issues that we encountered after training the network for a long time is that the resultant models would just output a constant value close to the mean of all sequences. Upon visual inspection we found that all the particles were converging to this value, as can be seen in fig. \ref{fig:mean-convergence-issue}. This means that either the transition function is ‘flattening’ and converting all pairs of input to the mean value, or that the observation likelihood function is giving low probability scores to every value except the mean of all the sequences. 

After further investigation we realized that because we have random input and because we’re using gradient descent certain behaviors such as tending to the mean are ‘encouraged’. This can be illustrated by the following example: imagine we have some stochastic process governed by a random walk, which centers around a mean of 0. We provide a notional example of such an equation below. Our goal is to train the neural network so that it models this equation in the transition function, given the input of the previous volatility and some random input. Now let's imagine we have a particle with a belief state of 0.5. Assuming our neural network is correctly pretrained the particle is equally as likely to move to 1 as it is to 0. However, because the stochastic process is centered around 0 the particle is likely to be penalized less if it moves to 0. Slowly the network learns not the transition function, but rather that regardless of the random input it should move the particle to the mean position (i.e. the position its least likely to be penalized in).

This results in the model very easily falling into a local minima in which the transition function simple transforms particles towards the mean. Alternatively the model could simply learn an observation probability function that assigns very high weights to particles close to the mean, thereby causing particle depletion everywhere else. To solve this we introduce a new loss function that weights the loss on each particle by the inverse likelihood of the belief state being at that position (given no prior information):

\begin{equation} \label{newlossfunc}
    \mathcal{L}(\theta) = \sum_{t=0}^{T} \sum_{i=0}^{N} \frac{\mathcal{L}(\theta|p_{t}^{i})}{f(p_{t+1}^{i})}
\end{equation}

In the case of our volatility model calculating \( f(p_{t+1}^{i}) \) is intractable, and so instead we use a gaussian distribution, with variance \( \tau \), and the mean \( \mu \) as an approximation.

\subsubsection{CUDA Compatibility}
In order to train the model faster the code was Cudified \cite{cuda}. CUDA is a framework that makes it easier to accelerate computation by enabling it to be run on parallel computing devices. In the case of our SV-PF-RNN training the neural networks can be massively sped up by training on a GPU. At first the size of the data, due to the length of the series and number of particles, caused the GPU to keep running out of memory. So we reduced the batch size to 50 and modified the code to clear the CUDA cache before running. Another side effect of running on a machine with a GPU is that random numbers may be generated slightly differently than they would be on a CPU \cite{PyTorchReproducability}. However, analyzing the effects of this was outside the scope of this work.

\section{Results and Evaluation}

In this section we evaluate the performance of our SV-PF-RNN, comparing it to the performance of a particle filter. We first go over the details of our experiment, and the metrics we use for comparison. We then present the results of these experiments. Finally we evaluate the performance of our particle filter, including an analysis on its performance with outliers and performance with different numbers of particles.

\subsection{Description of the Experiment}
\subsubsection{Evaluation Metrics}
For our evaluation we use mean squared error (MSE), mean absolute error (MAE), QLIKE, mean directional accuracy (MDA) and log likelihood:

\begin{equation} \label{MSE}
    MSE=\frac{1}{n}\sum_{t=0}^{T}(y_t-\hat{y}_t)^2
\end{equation}
\begin{equation} \label{MAE}
    MAE=\frac{1}{n}\sum_{t=0}^{T}\left | y_t-\hat{y}_t \right |
\end{equation}
\begin{equation} \label{QLIKE}
    QLIKE=\frac{1}{n}\sum_{t=0}^{T}(log(\hat{y}_{t}^{2})-\frac{y_{t}^{2}}{\hat{y}_{t}^{2}})
\end{equation}
\begin{equation} \label{MDA}
    MDA=\frac{1}{n}\sum_{t=1}^{T}1_{sign(\hat{y}_t-\hat{y}_{t-1})==sign(y_t-y_{t-1})}
\end{equation}
\begin{equation} \label{LogLikelihood}
    Log Likelihood=\sum_{t=0}^{T}log\left ( \frac{1}{\sqrt{2\pi\tau^2}}e^{-\frac{(\hat{y}_t-y_t)^2}{2\tau^2}} \right )
\end{equation}

\subsubsection{Experimental Setup}
For our experiment we generate \( K=3000 \) random paths of length \( T=300 \) using the stochastic volatility model described above. We fix our parameters at \( (\mu=-0.6558,\tau=0.1489,\phi=0.9807) \). For the SV-PF-RNN 2000 of these paths are used to train the network, 500 are used for testing and 500 are used for evaluation. 

\begin{figure}[h!]
\centering
\includegraphics[width = 0.8\textwidth]{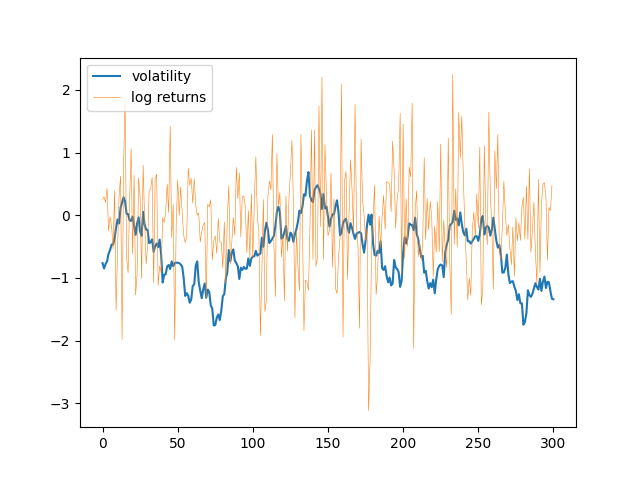}
\caption{Example of simulated volatility along with the simulated price innovations)}
\label{fig:logo}
\end{figure}

\subsection{Results}
\begin{table}[H]
\centering
\begin{tabular}{ |p{2.7cm}||p{2.6cm}|p{2.6cm}|p{2.6cm}|p{2.6cm}|  }
 \hline
 \multicolumn{5}{|c|}{Data Generated Using Taylor SV with \( 
(\mu=-0.6558,\tau=0.1489,\phi=0.9807) \)} \\
 \hline
 Model Type &Particle Filter&SV-PF-RNN (pretraining only)&SV-PF-RNN (standard loss function)&SV-PF-RNN (modified loss function)\\
 \hline
 MSE & 0.2210 & 0.2055 & 0.1823 & \textbf{0.1652}\\
 MAE & 0.3734 & 0.3637 & 0.3420 & \textbf{0.3246}\\
 QLIKE & 0.07510 & -1.7669 & -1.2038 & -0.8149\\
 MDA & 0.4970 & 0.4975 & 0.4946 & 0.4970\\
 Log Likelihood & -821 & -1094 & \textbf{-739} & -822\\
 \hline
\end{tabular}
\caption{Mean results from the experiment described in 5.2; best results are highlighted in bold (if significant)}
\end{table}

\subsection{Evaluation and Discussion}
After evaluating each of our models we find that our SV-PF-RNN with the modified loss function has the best performance out of all the models, followed by the SV-PF-RNN with the standard loss function. Interestingly the untrained SV-PF-RNN has a slightly lower MSE than the standard particle filter even though it is an approximation of the particle filter. We hypothesize this is due to the different resampling schema of the SV-PF-RNN, which prevents particles from depleting as quickly in low likelihood areas. Another interesting observation is that the log-likelihood of the particle filter is higher than the untrained SV-PF-RNN and around the same as the SV-PF-RNN with the modified loss function. The exact shape of the log-likelihood graph depends on the parameter \( \tau \). With \( \tau=0.1489 \) log-likelihood penalizes distant outliers more than MSE suggesting that the models with lower MSE may still suffer from having more distant outliers. 

\begin{figure}[htbp]
 \begin{minipage}{1\linewidth}
  \centering
  \includegraphics[width = 1 \linewidth]{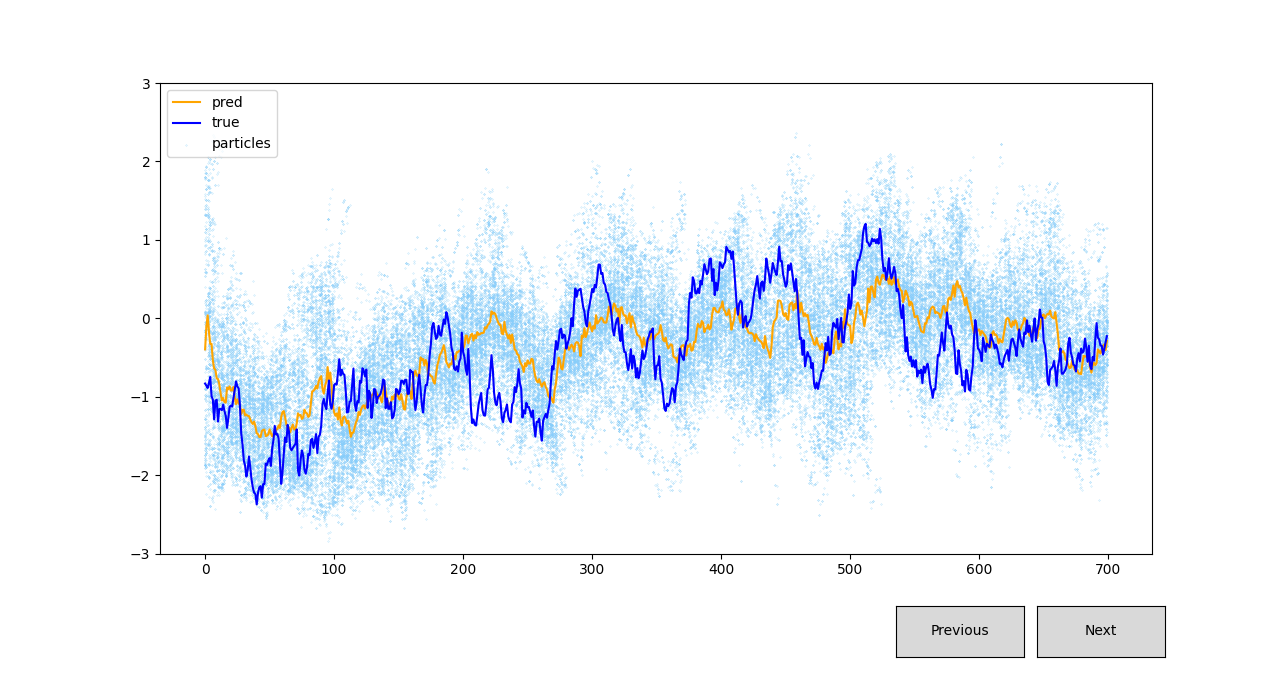}
  \caption{Pretraining Only}
  \label{fig:pretrained-model}
 \end{minipage}
 
 \begin{minipage}{1\linewidth}
  \centering
  \includegraphics[width = 1 \linewidth]{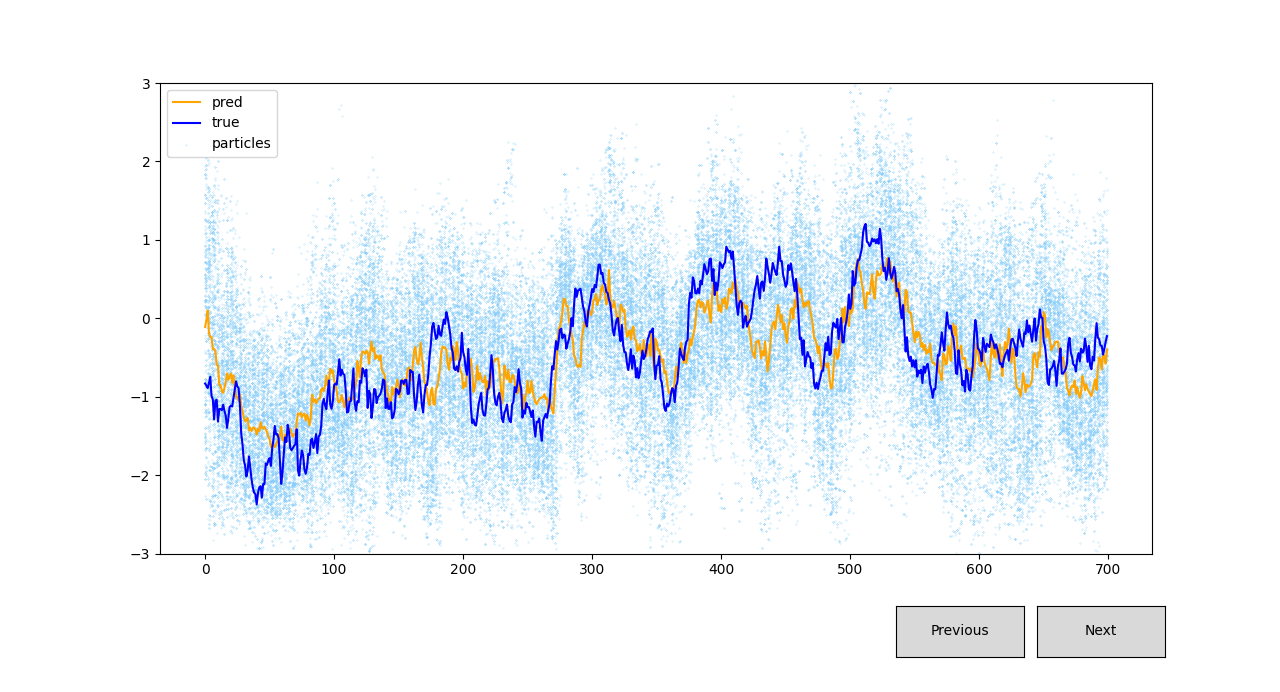}
  \caption{Fully Trained Model}
  \label{fig:fully-trained-model}
 \end{minipage}
\end{figure}

By plotting the particles and estimated volatility for our untrained and trained SV-PF-RNNs, figures \ref{fig:pretrained-model} and \ref{fig:fully-trained-model} respectively, we can visually see the difference between the two models. We can see that the SV-PF-RNN is a lot more responsive to sudden and large changes in volatility. Additionally the spread of particles is far larger. 

\subsubsection{Interpretability}
Another benefit of our SV-PF-RNN is the improved interpretability. By plotting the observation likelihood and transition functions we can see they’ve changed to get some insight into how the model has improved predictions. For real world data this could also give an insight into the estimated dynamics of a real world system. We have produced these two plots below. We can see that the observation likelihood has scored particles right next to the mean as lower than in the original (plotted in the contributions section). We can see the transition function has become a lot steeper, meaning that the spread of particles after the transition will be far larger. This is consistent with the larger particle spread seen in the graphs. Since we are generating data we know that these are not the true observation likelihood and transition functions, but rather may just improve the model’s performance as an estimator according to our loss function. 

\begin{figure}
\centering
\begin{minipage}{.5\textwidth}
  \centering
  \includegraphics[width=1\linewidth]{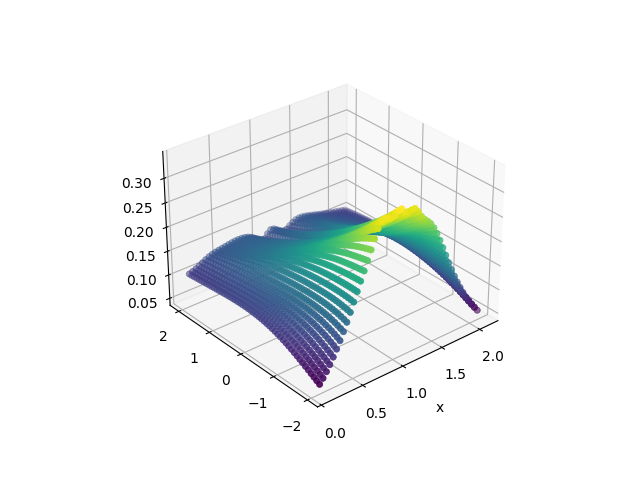}
  \captionof{figure}{Observation likelihood function in trained model}
  \label{fig:test1}
\end{minipage}%
\begin{minipage}{.5\textwidth}
  \centering
  \includegraphics[width=1\linewidth]{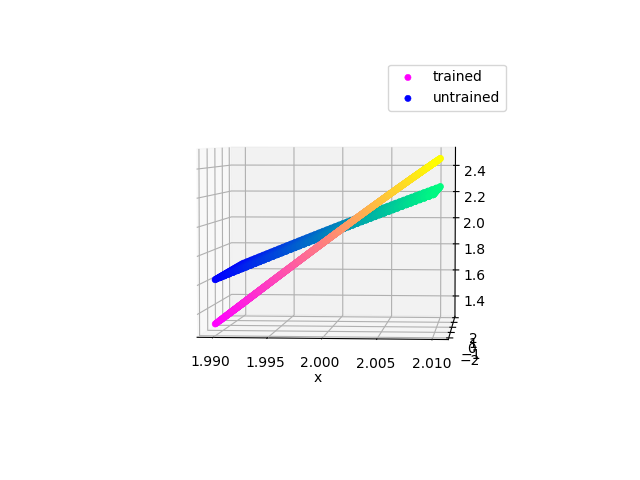}
  \captionof{figure}{Transition functions from trained and untrained models}
  \label{fig:test2}
\end{minipage}
\end{figure}

\newpage
\subsubsection{Performance on Outliers}

One of the disadvantages of machine learning methods over statistical methods is that they often perform badly with outliers in the data. To test our method against a regular particle filter we created a dataset in which the minimum or maximum values in each of the volatility series was in the 3rd or 97th percentile of all minimum or maximum values respectively. We then calculated the same set of statistics and plotted examples of the results.

\begin{table}
\centering
\begin{tabular}{ |p{2.7cm}||p{2.6cm}|p{2.6cm}|p{2.6cm}|  }
 \hline
 \multicolumn{4}{|c|}{Data Generated Using Taylor SV with \( 
(\mu=-0.6558,\tau=0.1489,\phi=0.9807) \)} \\
 \hline
 Model Type &Particle Filter&SV-PF-RNN (pretraining only)&SV-PF-RNN (fully trained)\\
 \hline
 MSE & \textbf{0.3313} & 0.5452 & 0.4654 \\
 MAE & \textbf{0.4527} & 0.5782 & 0.5363 \\
 \hline
\end{tabular}
\caption{Performance of the three models on outliers; best results are bolded}
\end{table}

\begin{figure}[h!]
 \begin{minipage}{1\linewidth}
  \centering
  \includegraphics[width = 1 \linewidth]{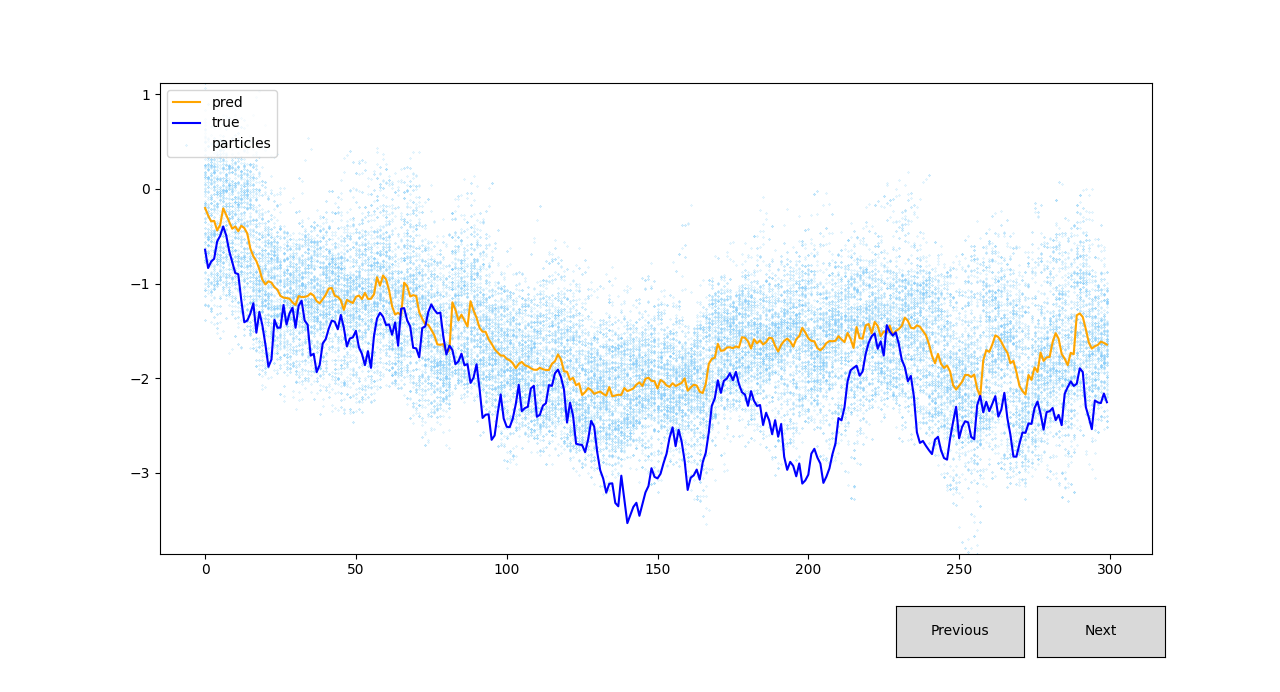}
  \caption{Particle Filter}
  \label{fig:particle-filter-outliers}
 \end{minipage}
 
 \begin{minipage}{1\linewidth}
  \centering
  \includegraphics[width = 1 \linewidth]{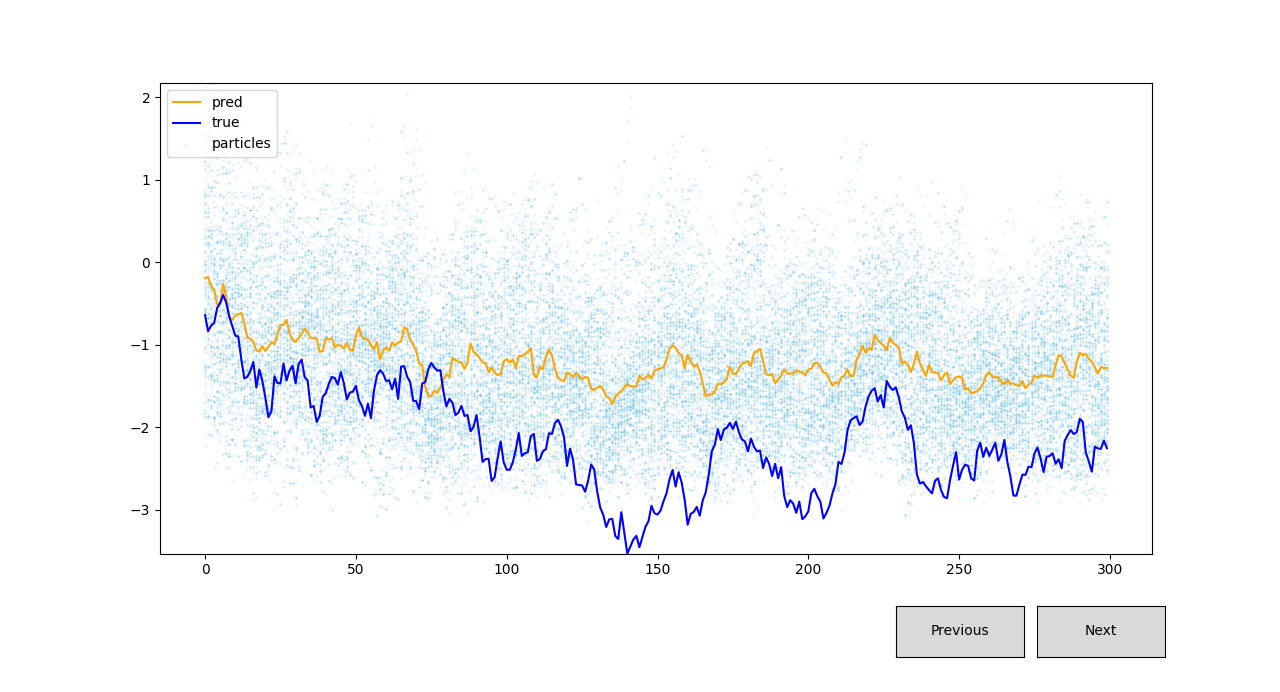}
  \caption{Fully Trained Model}
  \label{fig:sv-pf-rnn-outliers}
 \end{minipage}
\end{figure}

The model does not perform as well as the particle filter, with a 40.4\% increase in the mean squared error and an 18.5\% increase in the mean absolute error. The increase in the MSE is far more significant than the increase in the MAE suggesting that the PFRNN has far more regions in which it is significantly off from the true value of volatility. We can see in figures \ref{fig:particle-filter-outliers} and \ref{fig:sv-pf-rnn-outliers} that it often fails to converge to the true value of volatility for long periods of time.

Initially we hypothesized that is probably due to the fact that there were not as many extreme examples in the training data and so the model fails to generalize to these extreme cases. However the untrained model also performs even worse than the trained model, which suggests that the pretrained neural networks are simply not able to handle a large enough data range. There are two potential solutions to this problem: 
\begin{enumerate}
  \item Pretrain the network on a wider range of inputs (this may also require using larger networks).
  \item Balance the training dataset to include more examples of series with outliers.
\end{enumerate}

It is also worth noting the work of Pitt and Shephard, who observed that particle filters often poorly approximate the true tails of \( p(x_t|y_t) \), where \( x_t \) is the true state and \( y_t \) an observation \cite{Pitt1999}. This appears to be a weakness that our particle filter does not overcome. In their paper they adopt the auxiliary particle filter to mitigate the issues of particle degeneracy. By adapting our SV-PF-RNN to include auxiliary variables that encourage particle diversity we may also be able to improve performance on outliers. However, it should also be noted that this is not necessarily an issue with our SV-PF-RNN compared to the particle filter, as much as it is an issue with the type of particle filter we have used.

\subsubsection{Changing the Number of Particles}
\begin{table}
\centering
\begin{tabular}{ |p{2.6cm}||p{2.6cm}|p{2.6cm}||p{2.6cm}|p{2.6cm}| }
\hline
\multicolumn{5}{|c|}{Model Comparison} \\
\hline
\multicolumn{1}{|c|}{ } &
\multicolumn{2}{|c||}{Particle Filter} &
\multicolumn{2}{|c|}{SV-PF-RNN} \\
\hline
Number of Particles & MSE & MAE & MSE & MAE \\
\hline
16 & 0.6567 (0.3678) & 0.6422 (0.1826) & 0.4144 (0.1740) & 0.5149 (0.1045) \\
32 & 0.4932 (0.2266) & 0.5599 (0.1273) & 0.2858 (0.1019) & 0.4265 (0.0790) \\
64 & 0.3663 (0.1460) & 0.4824 (0.0991) & 0.2466 (0.0778) & 0.3985 (0.0642) \\
96 & 0.3369 (0.1381) & 0.4630 (0.0973) & 0.2393 (0.0736) & 0.3893 (0.0604) \\
128 & 0.3187 (0.1453) & 0.4459 (0.0926) & 0.2216 (0.0692) & 0.3756 (0.0588) \\
160 & 0.3119 (0.1253) & 0.4435 (0.0920) & 0.2323 (0.0742) & 0.3823 (0.0587)\\
\hline
\end{tabular}
\caption{Experiment Comparing the particle filter and SV-PF-RNN as we decrease the number of particles}
\label{table:particle-change}
\end{table}

Note: The variance values are shown in brackets below the respective MSE and MAE values.

\begin{figure}[h!]
\centering
\begin{minipage}{.5\textwidth}
  \centering
  \includegraphics[width=1\linewidth]{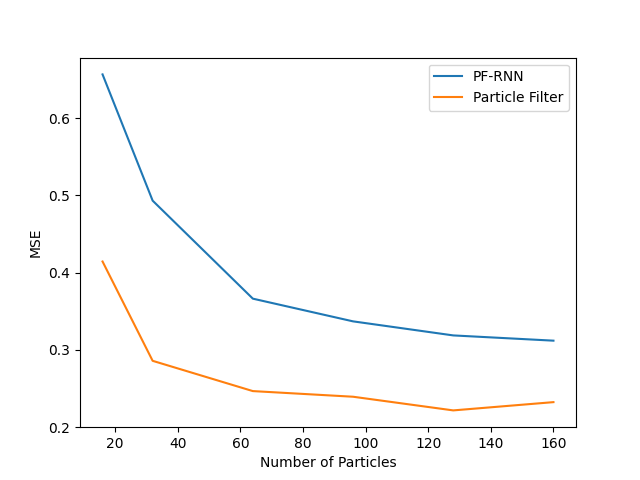}
  \captionof{figure}{MSE over Num. Particles}
  \label{fig:particles_vs_mse}
\end{minipage}%
\begin{minipage}{.5\textwidth}
  \centering
  \includegraphics[width=1\linewidth]{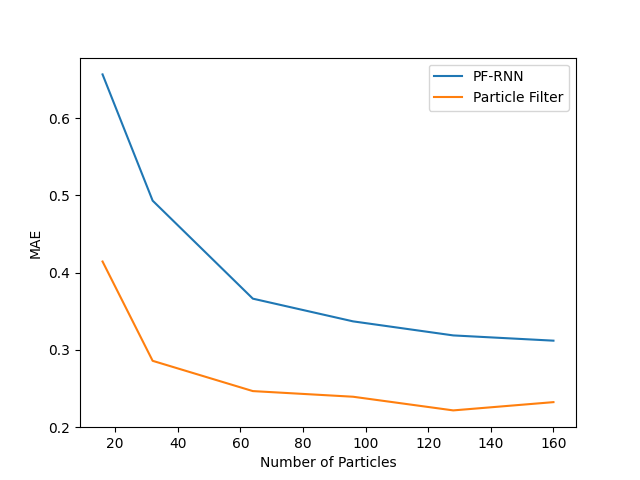}
  \captionof{figure}{MAE over Num. Particles}
  \label{fig:particles_vs_mae}
\end{minipage}
\end{figure}

Another interesting observation is how both the particle filters behave as the number of particles is increased or decreased. To test this we evaluated the trained PFRNN and the particle filter on 250 randomly generated paths with different numbers of particles. The results for this are shown in table \ref{table:particle-change}. Figures \ref{fig:particles_vs_mse} and \ref{fig:particles_vs_mae} show how the MAE and MSE changes for the two as we increase the number of particles.

We can see that for both filters the performance increases across the board as the number of particles is increased, with no significant performance increase going from 128 to 160 particles. It is also interesting to note that the variance of the results decreases as the number of particles increases. One interpretation of this is that the particle filter becomes more “stable” as the number of particles increases.

\section{Conclusion}

In this study, we proposed a novel hybrid architecture that combines a particle filter with a recurrent neural network (RNN) for volatility forecasting. Our objective was to enhance the performance of a particle filter and assess its effectiveness in both generated data and real-world scenarios.

First, we successfully improved upon the performance of a particle filter on the task of estimating volatility from generated data. Through extensive experimentation and evaluation, we observed notable enhancements in our hybrid model's accuracy and responsiveness to changes in volatility. This improvement signifies the potential of our approach to outperform traditional particle filters in capturing and forecasting volatility patterns.

Furthermore, we visually analyzed the generated data and plotted various graphs to investigate the behavior of our model. These visualizations demonstrated the increased responsiveness of our hybrid model to changes in volatility, reinforcing its ability to capture and adapt to dynamic market conditions.

However, when we applied our model to real-world data, we encountered challenges and achieved limited success. The nature of the real-world data differed significantly from the generated data used in our training process. The complex dynamics and unique characteristics of real-world volatility posed difficulties for our hybrid architecture, hindering its performance on this particular dataset.

\subsection{Future Work}
Despite the limitations encountered when training on real-world data, our study provides valuable insights into the capabilities and potential of hybrid particle filter-RNN models for volatility forecasting. Our findings suggest that further refinement and adaptation of our approach may be necessary to effectively tackle the intricacies and nuances present in real-world financial data.

\subsubsection{Incorporating Exogenous Variables}
One notable advantage of employing machine learning methods is their capacity to uncover intricate nonlinear patterns and relationships within data. By incorporating additional inputs after the pretraining phase, we can introduce exogenous explanatory variables that have the potential to enhance the model's ability to predict large volatility jumps. Previous research has demonstrated the effectiveness of including variables such as inflation rates or implied volatilities of related assets, leading to state-of-the-art performance using machine learning techniques \cite{Kim2018} \cite{Hu2020}.

\subsubsection{Adaptive Filtering}
A current limitation of our model is its assumption of static parameters governing volatility dynamics, despite evidence that parameters can change due to structural change in the market \cite{Liu2008}. The field of adaptive filtering offers promising avenues for improvement, involving continuous estimation of parameter values alongside volatility estimation \cite{Liu2001}. An alternative implementation of our model could involve incorporating the estimated parameter values as latent variables, with each particle having its own estimates of each parameter. A separate neural network could then be employed to iteratively update and adapt these values in real-time, resulting in a more dynamic and adaptive volatility forecasting framework.

\subsubsection{Implementing the Auxiliary Particle Filter}
One of the issues with our SV-PF-RNN was that it poorly estimated the tail ends of the distribution and thus failed to fit to outliers in the data. The auxiliary particle filter has been proposed as a solution to this problem \cite{Pitt1999}. Implementing this into our SV-PF-RNN, thereby creating the SV-APF-RNN, would have some benefits over a normal APF if successful. Usually implementing the APF would require the careful design of the function \( g(.) \). However, by implementing it as a differentiable network, the model can learn what this function should be.


\bibliographystyle{vancouver}

\end{document}